\documentclass[preprint,pre,aps]{revtex4}
\usepackage[dvips]{graphicx}
\usepackage{xcolor}
\usepackage{amsmath,amssymb}
\bibstyle{prsty}
\begin{document}
 \title{Mesoscopic inhomogeneities in concentrated electrolytes }
\author{O. Patsahan}
\affiliation{Institute for Condensed Matter Physics of the National Academy of Sciences of Ukraine, Lviv, Ukraine}
\author{  A. Ciach}
\affiliation{Institute of Physical Chemistry, Polish Academy of Sciences, 01-224 Warszawa, Poland}
 \date{\today} 
 \begin{abstract}
 	A mesoscopic theory for water-in-salt electrolytes combining density functional and field-theoretic methods is developed
 	in order to explain the unexpectedly large period of the oscillatory decay of the disjoining pressure  observed  in recent experiments for the LiTFSI (lithium bis(trifluoromethylsulfonyl)-imide) salt [T. S. Groves et. al., J. Phys. Chem. Lett. {\bf 12},
 	1702 (2021)]. We assumed spherical ions with different diameters, and implicit solvent inducing strong, short-range attraction between ions of the same sign. For this  highly simplified model, we calculated correlation functions. Our results indicate that mesoscopic inhomogeneities can occur when the sum of the Coulomb and the water-mediated interactions between like ions is attractive at short- and repulsive at large distances. We adjusted the attractive part of the potential to the water-in-LiTFSI electrolyte, and obtained both the period and the decay rate of the correlations in a semiquantitative agreement with the experiment. In particular, the decay length of the correlations increases nearly linearly with the volume fraction of ions.
 	\end{abstract}

\maketitle

\section{Introduction}
For many years it was commonly assumed that  dilute electrolytes, very well described by the Debye-H\"{u}ckel (DH) theory, are more suitable for electrochemical devices than the concentrated ones. For this reason,
 neither experimentalists nor theoreticians paid much attention to the concentrated electrolytes. 
  Recently, however, it was noted that the concentrated electrolytes have advantages such as a large electrochemical stability window and they may find applications in electrochemical devices, for example in lithium ion batteries~\cite{Chen2020,Suo2015,borodin:17:0,Groves2021}. These observations motivated intensive experimental studies. It was found that 
 when the concentration of ions, $\rho$, increases,  the deviation between experimental results and predictions of the DH theory becomes very large. Even qualitative trends well documented for the dilute electrolytes, such as decreasing screening length with increasing $\rho$, are opposite in concentrated electrolytes and ionic liquids (IL) solutions. Recent surface-force balance (SFB) experiments show that in the above systems, the screening length $\lambda_s$ is proportional to $\rho$, while the Debye screening length $\lambda_D$ perfectly describing the  dilute electrolytes is proportional to $1/\sqrt\rho$. The scaling behavior $\lambda_s/\lambda_D\sim (a/\lambda_D)^3$, where $a$ is the average diameter of the ions, was found for a number of concentrated solutions of simple salts in water, IL solutions and alkali halide solutions~\cite{Gebbie2013,smith:16:0,lee:17:0}.  This universal behavior suggests that the observed decay  of the disjoining pressure does not depend on specific interactions, but only on the Coulomb potential that is common for all the studied systems. 
 
 The strong disagreement of experimental results for concentrated electrolytes with the DH theory predictions attracted attention of theoreticians, but despite significant effort in theoretical and simulation studies, the experimental results are not fully explained yet~
 \cite{Kjellander2018,Kjellander2019,Goodwin2017,Ludwig2018,Rotenberg_2018,
 Adar2019,deSouza2020,Outhwaite2021,Coles2020,Zeman2020}. In several theories and simulation studies the scaling behavior   $\lambda_s/\lambda_D\sim (a/\lambda_D)^{\alpha}$ was found, but  the scaling exponent as well as $\lambda_s$ in these studies were significantly smaller than in the experiments~\cite{Goodwin2017,Ludwig2018,Rotenberg_2018,Adar2019,deSouza2020,Coles2020,Zeman2020}.
  Correct scaling for the charge-charge correlation length (that should be equal to $\lambda_s$) was obtained in \cite{ciach:21:0}, where it was shown that the local variance of the charge density plays a significant role for large concentrations of ions. However, oscillatory decay obtained in the theory is at variance with the asymptotic monotonic decay of the disjoining pressure observed in the experiments~\cite{Gebbie2013,smith:16:0,lee:17:0,Groves2021}. 
 
Theoretical studies of ionic systems are very often based on the restricted primitive model (RPM)~\cite{stell:95:0,fisher:94:0}. In the RPM, the ions are treated as charged hard spheres with the same diameter, and the Coulomb potential between the ions is assumed. The solvent, however, is treated as a dielectric continuum. The charge-charge correlations in the RPM decay monotonically in dilute systems, and exhibit exponentially damped oscillations for large $\rho$~\cite{leote:94:0,ciach:03:1}. More precisely, the oscillatory decay of charge-charge correlations occurs on the large-density side of the so called Kirkwood line~\cite{kirkwood:36:0} on the $(\rho, T)$ diagram. For the charge density profile near a planar electrode, as well as for the disjoining pressure between parallel planar electrodes,  the decay length and the period of oscillations are expected to be the same as the corresponding length in the charge-charge correlation function in the bulk electrolyte at the same thermodynamic state. The period of oscillations is close to $2a$~\cite{leote:94:0,ciach:03:1}, showing that the neighborhood of opposite charges is more probable than the neighborhood of like charges. The observed difference from the random distribution of charges is expected by the requirement of local charge-neutrality.   The predictions concerning the period of the damped oscillations were verified by simulations  and experiments~\cite{fedorov:08:0,fedorov:14:0,smith:16:0,lee:17:0,Zeman2020}. The decay length of the correlations in the RPM, however, depends on the approximation used in theoretical studies and remains a question of a debate~\cite{leote:94:0,patsahan:07:0,ciach:21:0}. In the SFB experiments, the oscillatory decay of the disjoining pressure was observed up to some distance between the crossed mica cylinders, but the asymptotic decay at larger distances was monotonic~\cite{Gebbie2013,smith:16:0,lee:17:0}. 

Strong disagreement with the RPM predictions was observed in recent SFB experiments for concentrated  LiTFSI (lithium bis(trifluoromethylsulfonyl)-imide)  in water~\cite{Groves2021}. It was found that the period of oscillations of the disjoining pressure was twice as large as the sum of diameters of the cation and the anion,  observed previously in many concentrated electrolytes and predicted by the RPM.
The same length scale of inhomogeneities   in the bulk  was  observed in scattering experiments for concentrated  LiTFSI~\cite{borodin:17:0}.  At large distances,  the decay of the disjoining pressure changes from oscillatory to monotonic, as found before for the other systems.
Importantly, the decay length  increases with increasing salt concentration and is of the same order of magnitude as observed previously in various concentrated electrolytes~\cite{Groves2021}.  

 In the LiTFSI  salt, the size and the chemical properties of the  $\rm{TFSI^-}$ and $\rm{Li^+}$ ions are significantly different~\cite{borodin:17:0}. The $\rm{TFSI^-}$ ion is not spherical, and is much larger than the $\rm{Li^+}$ ion. Moreover, the  $\rm{Li^+}$ ions are very well solvated in water, in contrast to the hydrophobic $\rm{TFSI^-}$ ions. 
Based on the above observations, we conclude that in the above water-in-salt electrolyte the size difference between the ions, and/or the specific  non-Coulombic interactions must play a significant role, and cannot be neglected.

The effect of specific interactions was studied in  the RPM supplemented with additional short-range (SR) interactions in Ref.\cite{ciach:01:0}. On the other hand, the size difference between the ions with neglected SR (primitive model, (PM)) was studied in Refs.\cite{ciach:07:0,patsahan2010gas,patsahan2011gas,patsahan:12:0}. It was found that the length scale of inhomogeneities depended on the strength of the SR interactions, and on the size asymmetry. These general predictions were not verified by experiments, however. In the particular case of  LiTFSI, the question of the origin of the scale of inhomogeneities and of the range of correlations is open.


In this work we  develop a minimal model for the water-in-salt electrolyte, and fit the parameters to the particular case of the  LiTFSI  salt. The minimal model can allow to see which details of the system properties can be neglected without changing the key features of the decay of the correlations.  The important questions are: (i) to what extent properties of the systems with large size asymmetry of ions and with strong specific interactions depend on details of the interactions and on the geometry of the ions? (ii) can the solvent be treated as a dielectric continuum that mediates effective ion-ion interactions, or must it be taken into account explicitly? (iii) what types of approximations should be used to compute the correlation functions  reproducing the qualitative trends found in experiments?

In our model we assume that both, the size difference and the effective SR interactions between the ions must be taken into account. The solvent, however, can be  treated as a dielectric continuum. We further assume that water induces effective ion-ion interactions of a range much shorter than the range of the Coulomb potential. The model is introduced in sec.\ref{model}.
In sec.\ref{theory}  we develop an approximate form of the grand-thermodynamic potential functional of local ionic densities that allows to obtain correlation functions.  In sec.\ref{results} we calculate the correlation functions  first in mean-field approximation (sec.\ref{MF}), and next beyond MF, using the procedure developed in \cite{ciach:20:1,patsahan:21:0} (sec.\ref{fluctuations}). We obtain a semiquantitative agreement with experiment when we assume strong water-induced short-range attraction between the cations. The last section contains our conclusions.

\section{Construction of the theoretical model}
\label{model}
In this section, we describe the assumptions and the approximations leading to the effective specific interactions between the  ions in  the  LiTFSI  salt dissolved in water. We take into account the ionic sizes reported in Ref.~\cite{Groves2021}. In addition, we develop the approximation for the SR interactions  based on the   requirement that the model predicts inhomogeneities at the length scale $\sim 2(\sigma_++\sigma_-)=4a $, where $\sigma_{\pm}$ denotes the diameter of the corresponding ion. Once the form of the SR interactions is assumed, we calculate the correlation functions for different volume fractions of ions  without further fitting of any parameters,  using the method summarized in sec.\ref{theory}. 

We first consider the excluded-volume interactions. 
The  $\rm{TFSI^-}$ ions are much bigger than the $\rm{Li^+}$ ions, and have a shape of an ellipsoid. We assume that the size difference is more important than the non-spherical shape, and assume that  the $\rm{Li^+}$ and  $\rm{TFSI^-}$ ions can be modeled as charged hard spheres with the diameter  $\sigma_+=0.2nm$ and $\sigma_-=0.6nm$, respectively. 

From the previous studies~\cite{borodin:17:0} we know that the  $\rm{Li^+}$ ions are strongly hydrophilic, while the  $\rm{TFSI^-}$ ions are strongly hydrophobic. The $\rm{Li^+}$  ions are solvated by water, and the  $\rm{TFSI^-}$ ions are not. This means that the short-range non-Coulombic forces have a strong effect on the distribution of the ions, and cannot be neglected. Even though the water- $\rm{Li^+}$ interactions play an important role, we assume that water can be treated as a dielectric continuum, as in the PM and the classical DH theory. We assume, however that water molecules mediate effective interactions between the  $\rm{Li^+}$ ions, and that these solvent-induced effective interactions are short-ranged (compared to the Coulomb potential) but strong. In addition, we assume short-ranged interactions between the  $\rm{TFSI^-}$ ions, but neglect the interactions between the  $\rm{Li^+}$ and  $\rm{TFSI^-}$ ions other than the Coulomb potential. We do not attempt to determine the precise shape of these interactions, because the correlations at large distances depend only on some gross features of the potentials, such as their range and strength. This is because in collective phenomena involving many  ions and solvent molecules,
many details are washed out by averaging over all distributions of the ions and the solvent molecules.

  With the above assumptions, we  model the aqueous electrolyte solution  as a binary mixture  of oppositely charged  hard spheres of different diameters ($\sigma_{+}\ne \sigma_{-}$) immersed in structureless dielectric medium with the dielectric constant $\varepsilon$.  We limit ourselves to the model with monovalent ions ($z_{+}=z_{-}=1$).
 The presence of the solvent is taken into account through the solvent-induced effective  short-range attractive interactions between the ions of the same sign. Therefore, we assume that the pair interaction potentials between two ions  for $r>\sigma_{\alpha\beta}=(\sigma_{\alpha}+\sigma_{\beta})/2$ , with $\alpha=+,-$ and $\beta=+,-$, 
 can be presented in the form:
\begin{equation}
U_{\alpha\beta}(r)= U_{\alpha\beta}^{C}(r)+U_{\alpha\beta}^{A}(r).
\label{u_alpha-beta}
\end{equation}
Here, $ U_{\alpha\beta}^{C}(r)$ are the Coulomb potentials between the ions with the signs $\alpha$ and $\beta$. As a length unit we choose the sum of radii, $a=(\sigma_{+}+\sigma_{-})/2$. The Coulomb potentials for $r^*\equiv r/a$ are
\begin{equation}
\label{U++}
\beta U_{++}^{C}(r^*)=\frac{l_B\theta(r^*-a_+)}{r^*},
\end{equation}
\begin{equation}
\label{U--}
\beta U_{--}^{C}(r^*)=\frac{l_B\theta(r^*-a_-)}{ r^*},
\end{equation}
\begin{equation}
\label{U+-}
\beta U_{+-}^{C}(r^*)=-\frac{l_B\theta(r^*-1)}{ r^*},
\end{equation}
where $a_{\pm}=\sigma_{\pm}/a$,  
\begin{equation}
\label{red_units}
l_B=\frac{1}{T^*}=\beta  E_{C}, \qquad E_{C}=\frac{e^2}{a\varepsilon},
\end{equation}
and  $\beta=1/k_{B}T$ with $k_{B}$ and $T$ denoting the Boltzmann constant and  temperature, respectively.  $l_B$ is the Bjerrum length in $a$ units, $E_C$ is the electrostatic potential of the pair of oppositely charged ions at contact, and the reduced temperature $T^*$ is in units of $E_C$. The unit step function $\theta(x)=1$ for $x>0$ and $\theta(x)=0$ for $x<0$,  prevents from contributions to the electrostatic energy of the pair of ions that would come from forbidden overlap of the hard cores. 

For $U_{\alpha\alpha}^{A}(r)$ we assume a short-range attractive potential.
The form of the sum of the direct (van der Waals type) and solvent-induced interactions is unknown, but we assume that its detailed shape is not necessary for studies of the collective phenomena such as the long-distance correlations. 
 For simplicity of calculations, we assume the attractive Yukawa potentials for $U_{\alpha\alpha}^{A}(r)$, 
\begin{equation}
\beta U_{\alpha\alpha}^{A}(r^*)=-l_B\epsilon^*_{\alpha\alpha}a_{\alpha}\displaystyle\frac{e^{-z^*_{\alpha}(r^*-a_{\alpha})}}{r^*}\theta(r^*-a_{\alpha}),\qquad
\alpha=+,-,
\label{Uii-r}
\end{equation} 
where $z^*_{\alpha}$ is in the $a^{-1}$ units, and we assume that $z^*_{+}=z^*_{-}=z=1.8$ to assure fast decay of these interactions. $\epsilon^*_{\alpha\alpha}$ measures the strength of the effective non-Coulombic interactions in units of $E_{C}$.
 Introducing the size-asymmetry parameter
\begin{equation}
\label{delta}
\delta=\frac{\sigma_--\sigma_+}{\sigma_-+\sigma_+}
\end{equation}
we get  $a_{\pm}=1\mp\delta$.

\begin{figure}[htbp]
	\begin{center}
		\includegraphics[width=0.45\textwidth,angle=0,clip=true]{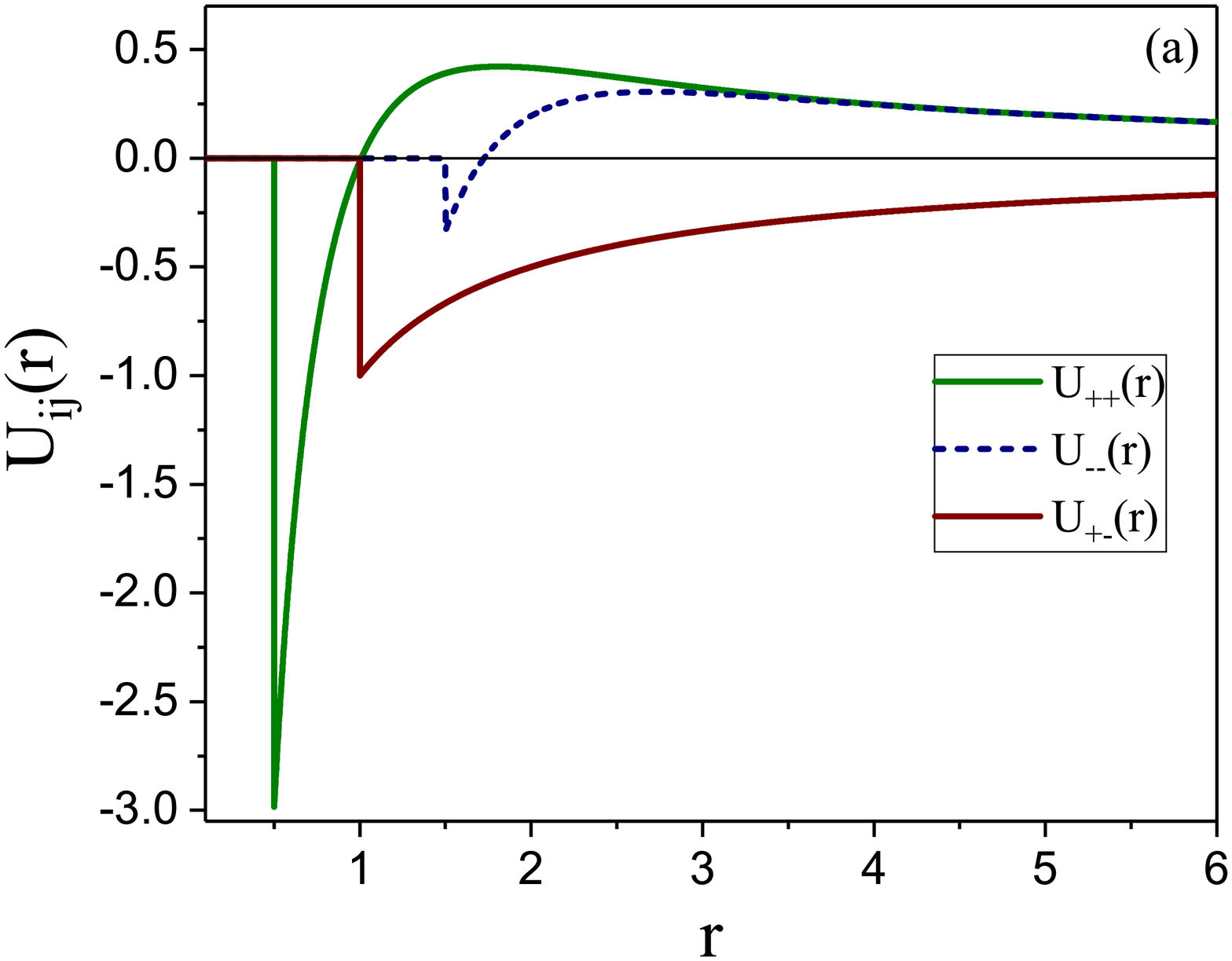}
		\includegraphics[width=0.45\textwidth,angle=0,clip=true]{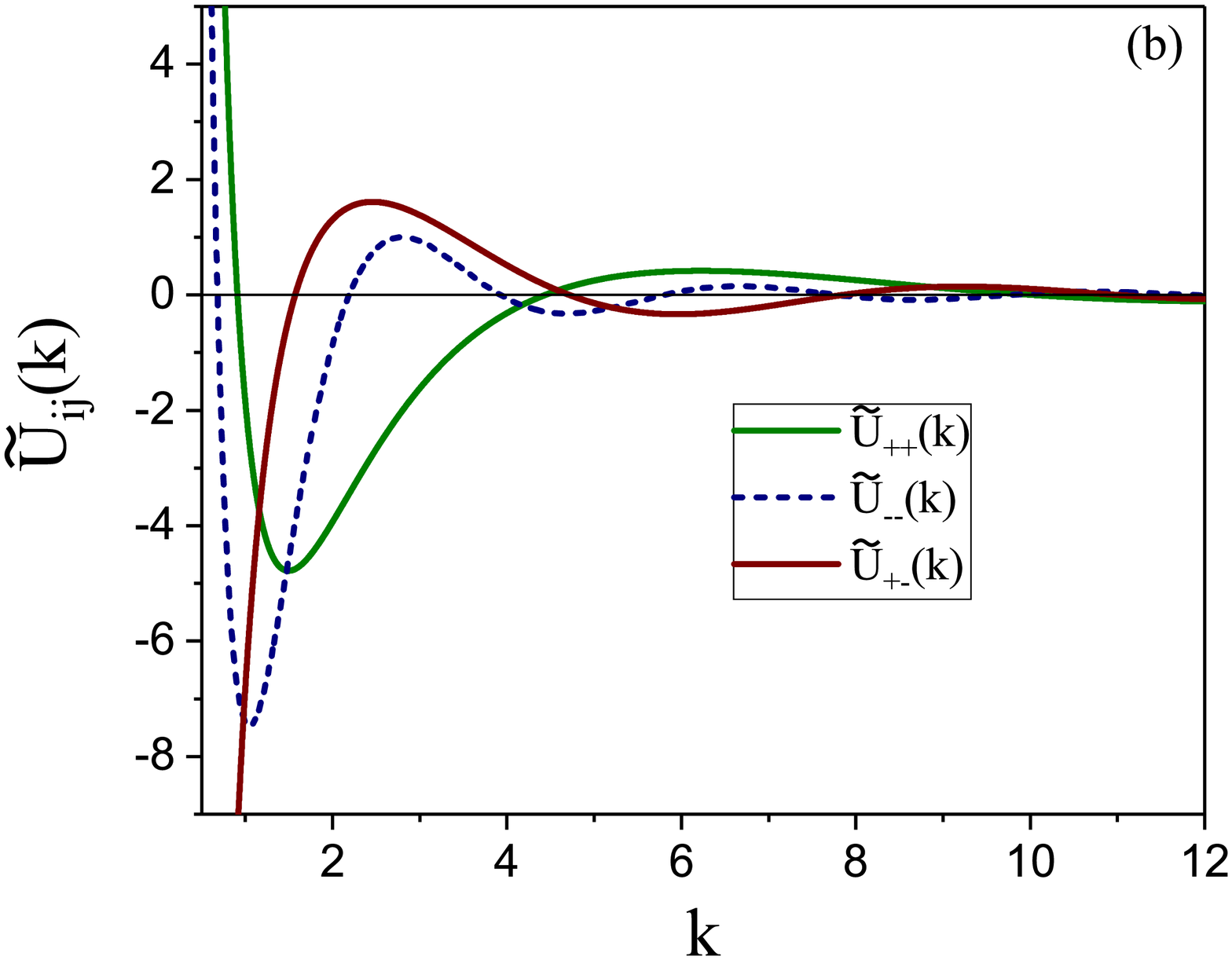}
		\caption{\label{fig1:Uij-r_PM-Y}  The interaction potentials $U_{\alpha\beta}(r)$  (Eqs.(\ref{u_alpha-beta})-(\ref{Uii-r})) for the model with $\delta=0.5$, $\epsilon_{++}^*=5$, and $\epsilon_{--}^*=1$ in real space (a) and in  Fourier representation (b).  $U_{\alpha\beta}$ are in units of $E_{C}$.
	 $E_{C}$ and  $\delta$  are defined in (\ref{red_units}) and (\ref{delta}).  $r$ and $k$ are in   $a$ and $a^{-1}$ units, respectively, where  $a=(\sigma_{+}+\sigma_{-})/2$.
		}
	\end{center}
\end{figure}

It remains to adjust the parameters $\epsilon^*_{\alpha\alpha}$ to  particular ions. 
We  assume that for the $\rm{TFSI^-}$ ions $\epsilon_{--}^{*}=1$, and treat  $\epsilon^{*}_{++}$ as a fitting parameter. In order to find the best approximation for the $\rm{Li^+}$ ions in water, we  calculate the length scale of inhomogeneities for arbitrary $\epsilon^{*}_{++}$ in sec.\ref{MF}.  
We find that 
  $\epsilon^{*}_{++}=5$ leads to a satisfactory agreement of the length scale of inhomogeneities  with  the experimental results. Based on this observation, we assume   $\epsilon^{*}_{++}=5$ for the considered system. Note that the second important length scale, $\lambda_s$, will be determined  for several concentrations of ions without additional fitting.

In Fig.~\ref{fig1:Uij-r_PM-Y}~(a)  the  potentials $U_{\alpha\beta}(r)$ normalized by $E_{C}=e^2/(a\varepsilon)$ are shown for the model with $\epsilon^{*}_{++}=5$,  $\epsilon_{--}^{*}=1$ and $\delta=0.5$. For the chosen parameters, the interaction potentials between  like ions consist of short-range attraction (SA)  and electrostatic long-range  repulsion (LR). Competing interaction potentials of this kind are also known  as SALR potentials.  In one-component systems,  the SALR-type  interactions can lead to spontaneously formed stable aggregates of particles,  such as spherical or elongated clusters, networks or layers~\cite{candia:06:0,archer:07:1,ciach:13:0}.
The Fourier transforms of the potentials $U_{\alpha\beta}(r)/E_{c}$ for the above mentioned model are shown in Fig. ~\ref{fig1:Uij-r_PM-Y}~(b), and the corresponding expressions are given in Appendix A. It is seen that 
$\tilde{U}_{++}(k)$ and $\tilde{U}_{--}(k)$
take minima at $k\neq 0$ which are rather close to each other.

\section{Theoretical formalism: a brief summary}
\label{theory}
In this section we present a brief description of the mesoscopic theory   for inhomogeneous mixtures \cite{ciach:11:2,ciach:20:1}.
In this  theory,  mesoscopic regions and mesoscopic states are considered. In a particular mesoscopic state, the volume fraction of ions with the $\alpha$ sign around the point ${\bf r}$,  $\zeta_{\alpha}({\bf r})$, is the fraction of the volume of the mesoscopic region occupied by this type of ions. The mesoscopic regions are comparable with or larger than $1$ (in $a$-units), and smaller than the scale of the inhomogeneities. The functions $\zeta_{\alpha}({\bf r})$ representing the  local   volume fraction can be considered as constraints imposed on the microscopic states. 
The local volume fraction of ions is given by $\zeta({\bf r})=\zeta_+({\bf r})+\zeta_-({\bf r})$. $\zeta({\bf r})$  averaged over the system volume is denoted by $\bar\zeta$.

The grand thermodynamic potential in the presence of the above mesoscopic constraints  can be written in the form
\begin{equation*}
\Omega_{co}[\zeta_{+},\zeta_{-}]=U_{co}[\zeta_{+},\zeta_{-}]-TS[\zeta_{+},\zeta_{-}]-\mu_{\alpha}\int d{\bf r}\zeta_{\alpha}({\bf r}),
\end{equation*}
 where $U_{co}$, $S$, and $\mu_{\alpha}$ are the internal energy, the entropy, and the chemical potential of the species $\alpha$, respectively. Hereafter, summation convention for repeated indices  is used. We make the approximation
 $-TS=\int d{\bf r} f_h(\zeta_{+}({\bf r}),\zeta_{-}({\bf r}))$, where $f_h(\zeta_{+}({\bf r}),\zeta_{-}({\bf r}))$ is the free-energy density 
 of the hard-core reference system in the local-density
 approximation,
 \begin{equation*}
 \beta f_{h}=\rho_{+}\ln\rho_{+}+\rho_{-}\ln\rho_{-}+\beta f_{hs},
 \end{equation*}
 where $f_{hs}$ is the contribution to the free energy density associated with packing of hard spheres with two different diameters. The expression for $f_{hs}$ in the Carnahan-Starling approximation  is given in Appendix B.

$U_{co}$ is given by the  expression 
\begin{equation*}
U_{co}[\zeta_{+},\zeta_{-}]=\frac{1}{2}\int_{\bf r_1}\int_{\bf r_2}
 V_{\alpha \beta}(|{\bf r}_1-{\bf r}_2|)\zeta_{\alpha}({\bf r}_1)\zeta_{\beta}({\bf r}_2).
\end{equation*}
 Because $\zeta_{\alpha}=\pi\rho_{\alpha}\sigma^3/6$ is used in the above definition, we have rescaled the interaction potential, 
 \begin{equation}
 \label{V}
 V_{\alpha\beta}(r)=\frac{U_{\alpha\beta}(r)}{v_{\alpha}v_{\beta}},
 \qquad
 v_{\alpha}=\pi\sigma^3/6.
 \end{equation}
 
 When the constraints imposed on the microscopic states by $\zeta_{+}({\bf r})$ and
 $\zeta_{-}({\bf r})$ are released, the microscopic states incompatible with $\zeta_{+}({\bf r})$ and
 $\zeta_{-}({\bf r})$ can appear, and the grand potential contains a fluctuation contribution and has the form~\cite{ciach:11:2}
 \begin{equation}
 \label{FF}
 \beta\Omega[\zeta_+,\zeta_-]=\beta\Omega_{co}[\zeta_+,\zeta_-]-
 \ln\Big[\int D\phi_{+}  \int D\phi_{-} e^{-\beta H_{fluc}}\Big],
 \end{equation}
 where 
 \[
 \label{Hfl}
 H_{fluc}=\Omega_{co}[\zeta_++\phi_+,\zeta_-+\phi_-]-\Omega_{co}[\zeta_+,\zeta_-]
 \] 
 is associated with the appearance of the fluctuation $\phi_{\alpha}$ of  the local volume fraction $\zeta_{\alpha}$. In MF, the second term on the RHS of Eq.(\ref{FF}) is neglected.
 
We are interested in the correlation functions  in the disordered phase
 \begin{eqnarray}
 \label{G}
 G_{\alpha\beta}({\bf r})=\langle \Delta\zeta_{\alpha}({\bf r}_0) \Delta\zeta_{\beta}({\bf r}+{\bf r}_0)\rangle,
 \qquad
\alpha,\beta=+,-
 \end{eqnarray}
 where $\Delta \zeta_{\alpha}({\bf r})=\zeta_{\alpha}({\bf r})-\bar\zeta_{\alpha}$, and
 $\bar\zeta_{\alpha}$ is the average volume fraction  of the ions with the $\alpha$ sign.
The matrix ${\bf G}$ with the elements
defined in (\ref{G}) satisfies the analog of the Ornstein-Zernicke equation, ${\bf G}={\bf C}^{-1}$, where the inverse correlation 
functions  $\tilde C_{\alpha\beta}$ are the
second functional derivatives of $ \beta\Omega[\zeta_+,\zeta_-]$ with respect to $\zeta_{\alpha}$ and $\zeta_{\beta}$~\cite{ciach:11:2}.  

In the lowest-order nontrivial approximation beyond MF~\cite{ciach:11:2,ciach:20:1},
\begin{eqnarray}
\label{C}
\tilde C_{\alpha\beta}(k)=\beta \tilde V_{\alpha\beta}(k)+A_{\alpha\beta}+\frac{A_{\alpha\beta\gamma\delta}}{2}{\cal G}_{\gamma\delta},
\end{eqnarray}
where $\tilde f(k)$ denotes the function $f$ in Fourier representation. In the above equation,
\begin{eqnarray}
\label{A4}
A_{\alpha_1....\alpha_j}=\frac{\partial^j \beta f_h( \zeta_+, \zeta_-)}
{\partial\zeta_{\alpha_1}...\partial\zeta_{\alpha_j}},
\end{eqnarray}
with $\alpha_i=+,-$.
Note that in this approximation, the dependence of $\tilde C_{\alpha\beta}(k)$ on $k$ comes only from $\beta \tilde V_{\alpha\beta}(k)$.
The last term in Eq.(\ref{C}) is the  fluctuation contribution, 
and comes from the last term in (\ref{FF}) in the Brazovskii-type approximation~\cite{brazovskii:75:0}.
Here, ${\cal G}_{\gamma\delta}$ 
denotes the integral
\begin{eqnarray}
\label{calG}
{\cal G}_{\gamma\delta}=\int\frac{d{\bf k}}{(2\pi)^3}\tilde G_{\gamma\delta}(k).
\end{eqnarray}
Eqs.(\ref{C})-(\ref{calG}) have to be solved self-consistently. In general, it is a nontrivial task.

Note that ${\cal G}_{\alpha\alpha}=\langle \Delta\zeta_{\alpha}({\bf r}) \Delta\zeta_{\alpha}({\bf r})\rangle$ is a local variance of $\zeta_{\alpha}$, i.e. $\sqrt{\cal G}_{\alpha\alpha}$ is the standard deviation  from the space-averaged value of the local volume fraction of the $\alpha$-ions. The larger is  ${\cal G}_{\alpha\alpha}$, the stronger are the mesoscopic inhomogeneities. 

We focus on the disordered inhomogeneous phase and  assume that the inhomogeneities 
occur on a well-defined length scale. In such a case, the peak of $\tilde G_{\gamma\delta}(k)$ (proportional to the structure factor) is high and narrow. 
For functions with a high, narrow peak, the main contribution to the integral comes 
from the vicinity of the maximum.  
We assume that the maximum of all the integrands
in (\ref{calG}) is very close to the minimum  at $k=k_0$ of $\det \tilde {\bf C}(k)$,
and we make the approximation
\begin{equation}
\label{calG2}
{\cal G}_{\alpha\beta}=[ \tilde C_{\alpha\beta}(k_0)]{\cal G},
\end{equation}
where $[ \tilde C_{\alpha\alpha}(k)]=\tilde C_{\beta\beta}(k)$ and
$[ \tilde C_{\alpha\beta}(k)]=-\tilde C_{\alpha\beta}(k)$ for $\alpha\ne\beta$, and
\begin{eqnarray}
\label{calG3}
{\cal G}=\int\frac{d{\bf k}}{(2\pi)^3}\frac{1}{\det \tilde {\bf C}(k)}.
\end{eqnarray}
Near the minimum at $k_0$, we have the approximation
\begin{eqnarray}
\label{detC}
\det \tilde {\bf C}(k)= D_0+\frac{\beta\tilde W^{''}(k_0)}{2}(k-k_0)^2+...
\end{eqnarray}
where 
\begin{eqnarray}
\label{D0}
D_0
=\det\tilde {\bf C}(k_0),
\end{eqnarray}
and 
$\beta W^{''}(k_0)$ is the second-order derivative of $\det\tilde {\bf C}(k)$ with respect to the wave number $k$ at $k=k_0$.
From the approximation (\ref{detC}) and (\ref{calG3}), we obtain~\cite{ciach:11:2,ciach:12:0}
\begin{equation*}
{\cal G}\approx
\frac{k_0^2}{\pi\sqrt{2\beta \tilde W^{''}(k_0)D_0}}.
\end{equation*}

With all the above assumptions, the problem reduces to determination of the minimum of $\det \tilde {\bf C}(k)$,
and to a solution of  three algebraic equations for $ \tilde C_{\alpha\beta}(k_0)$ (see (\ref{C}) and (\ref{calG2})) because,
\begin{eqnarray}
\label{C(k)}
\tilde C_{\alpha\beta}(k)=  \tilde C_{\alpha\beta}(k_0) +\beta\Delta\tilde V_{\alpha\beta}(k),
\quad
\alpha,\beta=+,-
\end{eqnarray}
 where
 \begin{equation}
 \label{Delta_V}
\Delta\tilde V_{\alpha\beta}(k) =\tilde V_{\alpha\beta}(k)-\tilde V_{\alpha\beta}(k_0)\approx \frac{V_{\alpha\beta}^{''}(k_0)}{8k_0^2}(k^2-k_0^2)^2.
 \end{equation}
The last approximation is valid for  $k\approx k_0$.

It should be noted that the results obtained within the framework of this theory  for several models of inhomogeneos mixtures were verified by simulations \cite{ciach:20:1,patsahan:21:0,Patsahan2021}. 

\section{Results}
\label{results}
\subsection{MF approximation}
\label{MF}
In MF, we neglect the last term in Eq.~(\ref{C}), and easily obtain explicit expressions for the matrix $\tilde {\bf C}^{MF}(k)$ inverse to the matrix of correlations. These expressions are  shown in Appendix C. 
In MF, the disordered phase becomes unstable with respect to oscillatory  modulations of the volume fractions of the ions at the so-called  $\lambda$-line on the $(\bar\zeta,T^*)$ diagram. The $\lambda$-line marks the boundary of stability of
the disordered phase with respect to mesoscopic fluctuations of the volume fractions and separates
the phase space into regions corresponding to the homogeneous and inhomogeneous (on the
mesoscopic length scale) phases. Thus, in MF the $\lambda$-line is interpreted as a continuous order-disorder transition.
In order to get the $\lambda$-line, one should solve the system of equations
\begin{eqnarray*}
& \det\tilde {\bf C}^{MF}(k_0)=0, \nonumber \\
& \left. \displaystyle\frac{\rm{d}\det\tilde {\bf C}^{MF}}{\rm{d}k}\right|_{k=k_0} =0.
\end{eqnarray*}

When the $\lambda$-line is crossed, the  inhomogeneities in the distribution of ions  occur on the length scale $2\pi/k_0$. In our model, $k_0$ depends in particular on $\epsilon^*_{++}$ that we left as a free parameter. In order to fit $\epsilon^*_{++}$ to our water-in-salt system, we need to have $2\pi/k_0\approx 4$ in $a$  units ($k_0\approx 1.6$ in  $a^{-1}$ units),  for the molarity $M\sim 3-5$ for which the experimental data were obtained. The volume fraction of the spherical ions with $\sigma_+=0.2nm$ and $\sigma_-=0.6nm$ is related to the  molarity $M$ by $\bar\zeta/M= \frac{\pi}{6}(0.2^3+0.6^3)10^{-24}N_A $, where $N_A$ is the Avogadro number. We get $\bar\zeta\approx 0.27$ and $\bar\zeta\approx 0.32$ for the  $3.8M$ and $4.6M$ systems, respectively. However, the above formula is a very rough estimation for $\bar\zeta/M$ in view of the strong dependence of $\bar\zeta$ on the diameter of the ions and the ellipsoidal shape of the  $\rm{TFSI^-}$ anions, and it only gives the order of magnitude of $M$ in the experimental system 
for given $\bar\zeta$ in our theory.
Thus, in our semiquantitative analysis, we will consider volume fractions up to $\bar \zeta=0.55$.
  
\begin{figure}[htbp]
	\centering
		\includegraphics[width=0.45\textwidth,angle=0,clip=true]{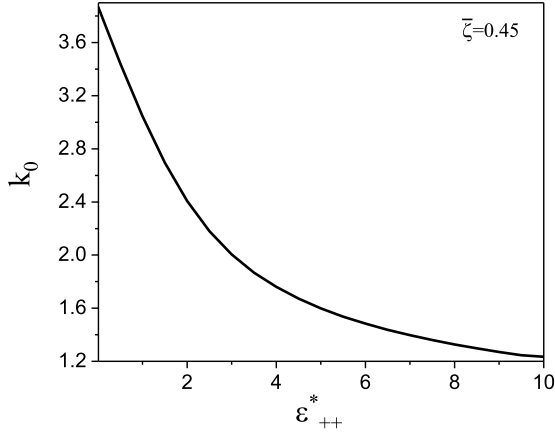}
\caption{\label{fig:k0}{The wavenumber of the density waves $k_0$ (in $a^{-1}$ units) as a function of $\epsilon^*_{++}$ for  $\epsilon^*_{--}=1, \delta=0.5$ and the volume fraction of ions $\bar\zeta=0.45$. $\epsilon^*_{\alpha\alpha}$ describes the strength of the non-Coulombic interactions (see (\ref{Uii-r})), and the size asymmetry $\delta$ is defined in (\ref{delta}). For the considered system, $a=(\sigma_{+}+\sigma_{-})/2\approx 0.4nm$.}
		}
\end{figure}

The plot of $k_0$ as a function of $\epsilon^*_{++}$ for $\epsilon^*_{--}=1, \delta=0.5$ and $\bar\zeta=0.45$ is shown in Fig.\ref{fig:k0}.
We can see that for $\epsilon^*_{++}=5$, the length scale of inhomogeneities is $2\pi/k_0\approx 3.9$ (in $a$-units), which for $a=0.4nm$ gives $1.56nm$ that is close to the experimental result $1.4nm$. We thus choose $\epsilon^*_{++}=5$ in our further calculations.
\begin{figure}[htbp]
	\centering
		\includegraphics[width=0.45\textwidth,angle=0,clip=true]{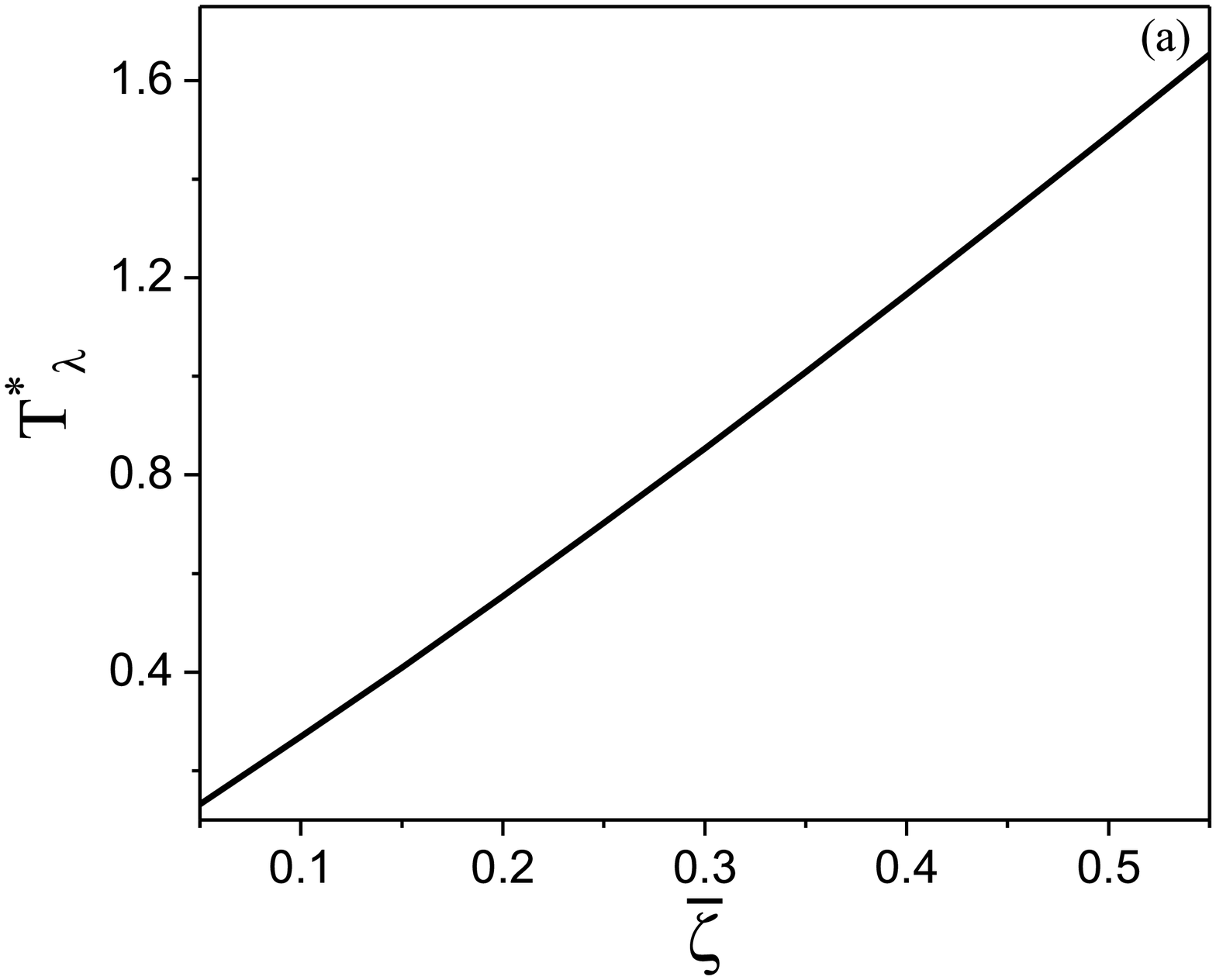}
		\includegraphics[width=0.45\textwidth,angle=0,clip=true]{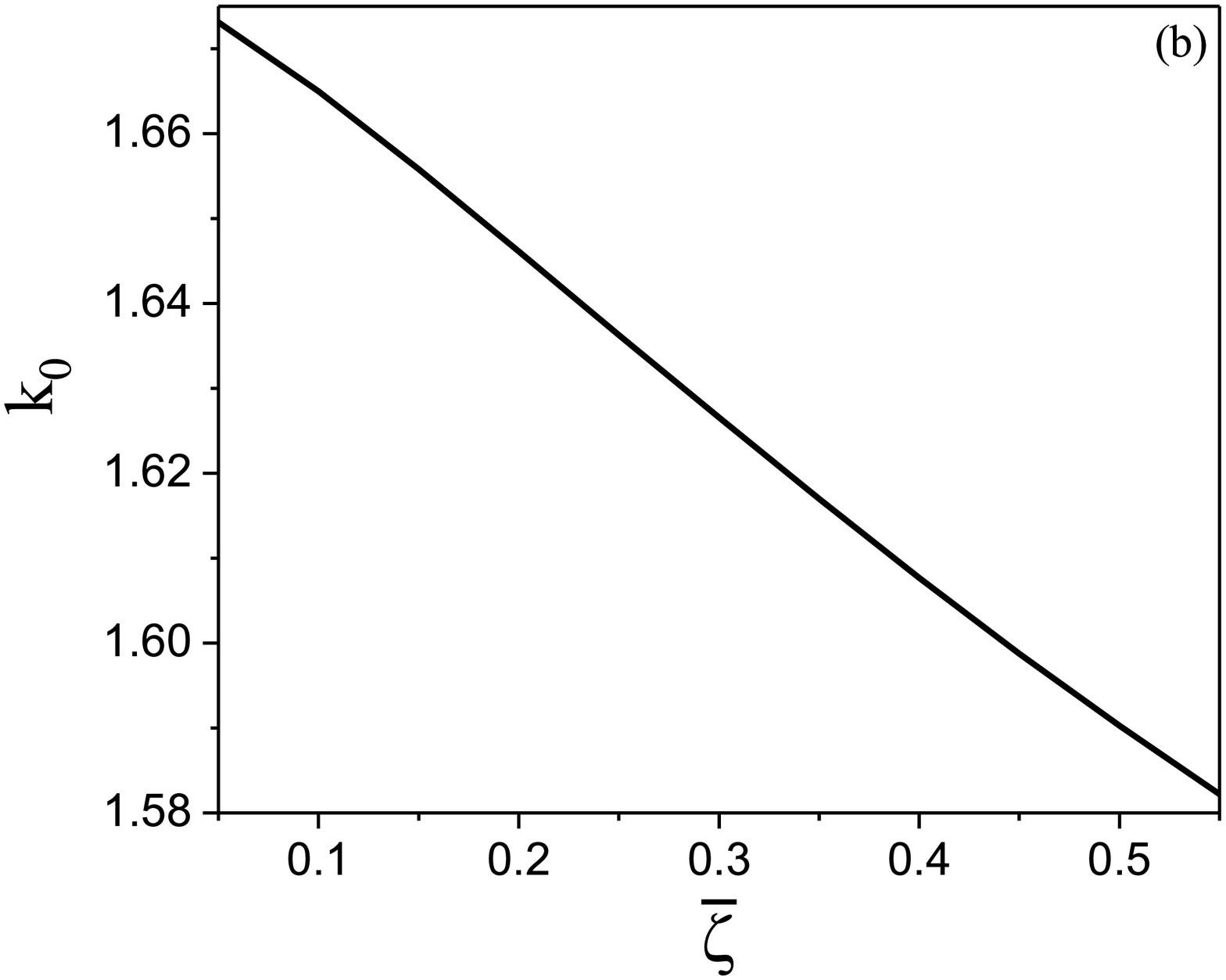}
		\caption{\label{fig2:lambda} {$\lambda$-line in $\bar\zeta$---$T^*$ (panel~a) and $\bar\zeta$---$k_{0}$ (panel~b) coordinates for the model with $\delta=0.5$,  $\epsilon_{++}^*=5.0$, and $\epsilon_{--}^*=1.0$.  
		$T^{*}$  and $\delta$ are defined in (\ref{red_units}) and (\ref{delta}), respectively. $\bar\zeta=\bar\zeta_{+}+\bar\zeta_{-}$ is the total volume fraction of ions,  $\bar\zeta_{\alpha}=\pi\rho_{\alpha}\sigma_{\alpha}^3/6$, $k_{0}$ is in the $a^{-1}$ units, $a=(\sigma_{+}+\sigma_{-})/2\approx 0.4nm$.}
		}
\end{figure}

Fig.~\ref{fig2:lambda} shows the  $\lambda$-line in the $\bar\zeta$---$T^*$ (panel~a) and $\bar\zeta$---$k_{0}$ (panel~b) coordinates for the model with $\delta=0.5$, $\epsilon_{++}^*=5$, and $\epsilon_{--}^*=1$.  
For the thermodynamic states  below the $\lambda$-line, 
the waves with the wavelength $2\pi/k_{0}$ are more probable than the constant volume fractions. For our model, the emergence of the inhomogeneous structure for $ \det \tilde {\bf C}^{MF}(k_0)<0$ may be associated with the formation of aggregates such as  clusters or layers, rather than with a phase transition.  For such thermodynamic
states the fluctuations dominating  on the mesoscopic length scale
 should be taken into account in order to restore the stability of the disordered phase. As found in different systems with spontaneously appearing mesoscopic inhomogeneities, fluctuations induce a change of the continuous transition found in MF to the first-order crystallization that is also shifted to higher volume fractions. 
 Because of the instability of the disordered phase for $T^*<T_{\lambda}^{*}$ in MF,  the asymptotic decay of the correlation functions $G_{\alpha\beta}^{MF}(r)$ can be analyzed only for $T^*>T_{\lambda}^{*}$.
 
  In general,  the long-range behaviour  ($r\gg 1$) of $G_{\alpha\beta}(r)$  is described by the function~\cite{evans:94:0}
 \begin{equation}
 \label{Gr}
 G_{\alpha\beta}(r)={\cal A}_{\alpha\beta}e^{-\alpha_0 r}\sin(\alpha_1 r+\theta_{\alpha\beta})/r.
 \end{equation}
 In (\ref{Gr}),  $\alpha_{0}=1/\lambda_s$ and  $\alpha_{1}=2\pi/\lambda$ are the imaginary and  real  parts of the leading order pole of 
 $\tilde G_{\alpha\beta}(q)$ in the complex $q$-plane, which is determined as the complex root  $q =i\alpha_0 \pm\alpha_1$  of  the equation  $ \det \tilde {\bf C}(q)=0$  having the smallest imaginary part. Since all $\tilde G_{\alpha\beta}(q)$  have a common denominator  $ \det \tilde {\bf C}(q)$, they  exhibit the same pole structure and have the same
 exponential contributions. Only the amplitudes ${\cal A}_{\alpha\beta}$  and the phases $\theta_{\alpha\beta}$ differ  for different $\alpha\beta$ combinations. 
 \begin{figure}[h]
 	\includegraphics[clip,width=0.6\textwidth,angle=0]{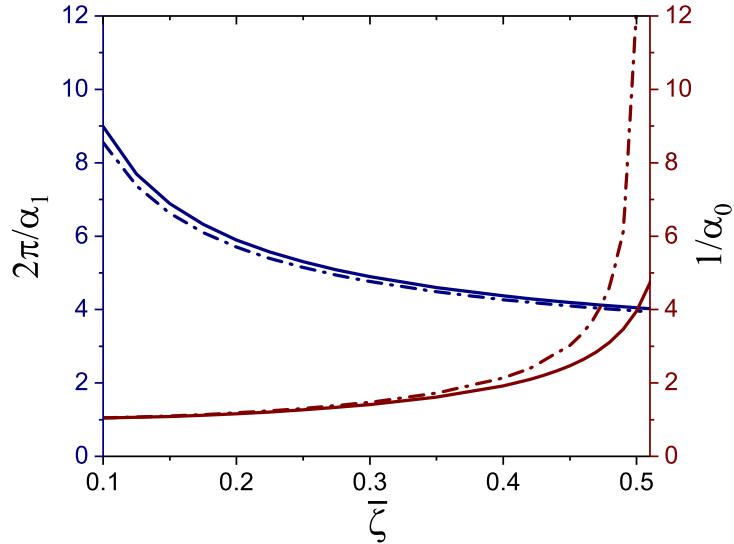}
 	\caption{\label{fig3:alpha}
 		(Colour online)	 The model with $\delta=0.5$,  $\epsilon_{++}^*=5.0$, and $\epsilon_{--}^*=1.0$. The decay length $\lambda_s=\alpha_{0}^{-1}$ and the period of oscillations $\lambda=2\pi/\alpha_{1}$ of the correlation functions $G_{\alpha\beta}(r)$ as functions of the total volume fraction of ions $\bar\zeta$ for $T^*=1.5$ (solid line) and $T^*=1.6$ (dash-dotted line) in the MF approximation.  $\alpha_{0}$ and $\alpha_{1}$ are in the $a^{-1}$ units, with $a=(\sigma_{+}+\sigma_{-})/2\approx 0.4nm$.
 	}
 \end{figure}

 We calculate  $\alpha_{0}$ and $\alpha_{1}$   for our model in MF from the equation  $ \det \tilde {\bf C}^{MF}(q)=0$. The $\bar\zeta$-dependence of both, the decay length $\lambda_s=\alpha_{0}^{-1}$ and the period of oscillations $\lambda=2\pi/\alpha_{1}$ of the correlation functions $G_{\alpha\beta}(r)$ is presented in Fig.~\ref{fig3:alpha}  for two values of the reduced temperature, $T^*=1.5$ and $T^*=1.6$ ($T^*>T_{\lambda}^*$). Note  that the decay length $\alpha_{0}^{-1}$ tends to $\infty$ ($\alpha_{0}\to 0$)  and simultaneously  the period of oscillations $\lambda$ tends to $2\pi/k_0\approx 4a$ when $T^*\to T_{\lambda}^*$. More precisely, we get $\lambda=3.96\,a$ for $T^*=1.5$, $\bar\zeta=0.5$ and  $\lambda=3.98\,a$ for $T^*=1.6$, $\bar\zeta=0.55$.

 There is a very small difference between the values of $\lambda$ obtained for the two temperatures and the difference decreases with an increase of $\bar\zeta$.  
 Moreover,  for $\bar\zeta>0.3$ the dependence of $\lambda$ on $\bar\zeta$ is weak, as in \cite{Groves2021}.
 
The values of the reduced temperature $T^*>T_{\lambda}^*$ for large $\bar\zeta$,   however,  are too high when compared to room temperature. A rough estimate of the reduced temperature that corresponds to the conditions for the LiTFSI salt in water at room temperature  ($T=300^\circ$C and $\varepsilon=80$) is about $0.5$. Assuming
  that the dielectric constant of bulk water is decreased proportionally to the ion concentration, we should consider $T^*<0.5$.

 \subsection{Beyond the MF approximation}
 \label{fluctuations}
 In order to calculate the fluctuation contribution to the inverse correlation functions $\tilde C_{\alpha\beta}(k)$ in the Brazovskii-type approximation, we take into account the last term in (\ref{C}) and solve  the closed set of four equations for the unknowns  $k_0$ and $\tilde C_{\alpha\beta}(k_0)$. The explicit forms of these equations are given in Appendix D.
Once $k_0$ and $\tilde C_{\alpha\beta}(k_0)$ are determined, the inverse correlation functions $\tilde{C}_{\alpha\beta}(k)$ can be obtained from Eqs.~(\ref{C(k)})--(\ref{Delta_V}).

\begin{figure}[htbp]
	\begin{center}
		\includegraphics[width=0.45\textwidth,angle=0,clip=true]{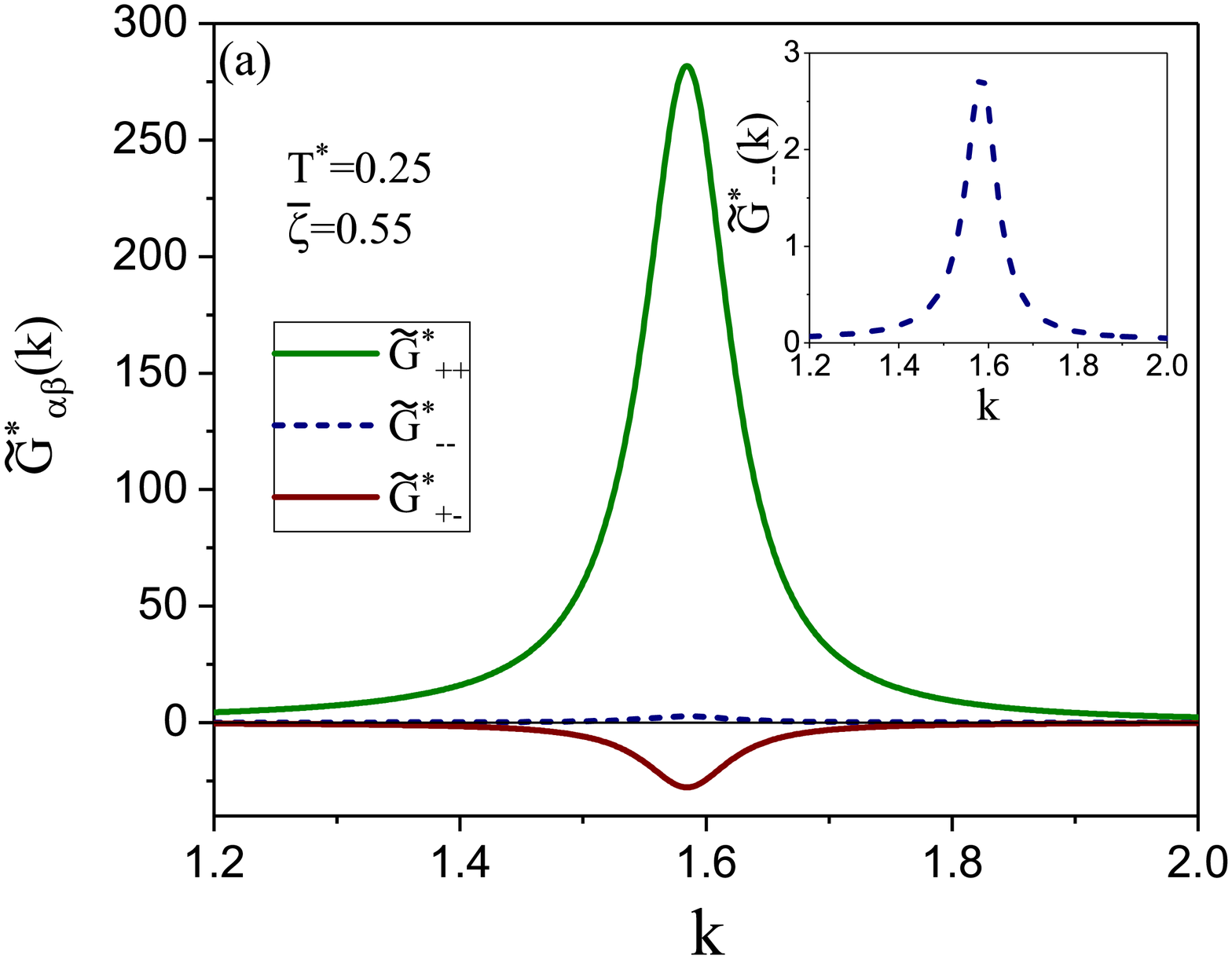}
		\includegraphics[width=0.46\textwidth,angle=0,clip=true]{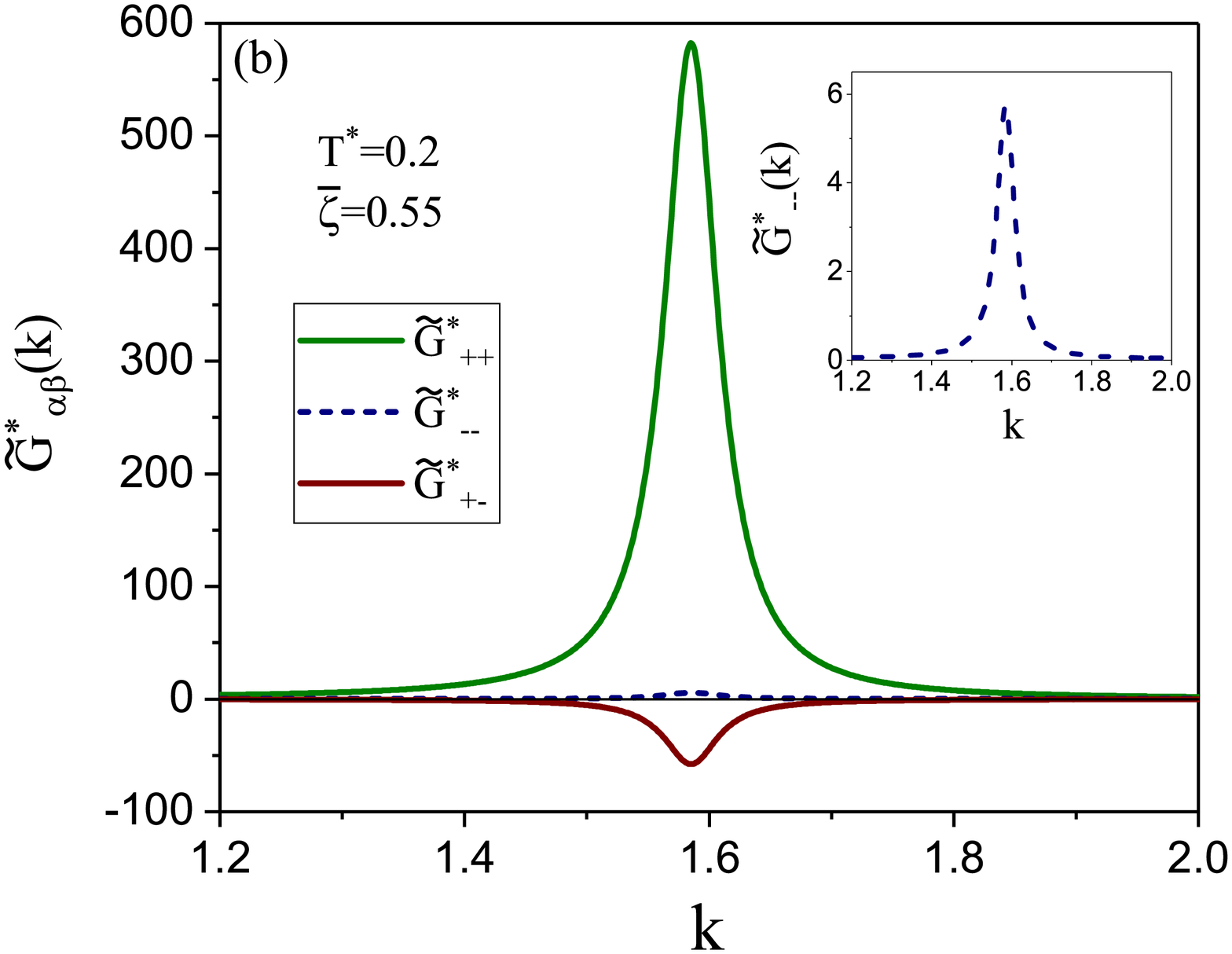}
		\caption{\label{fig:Gij_fl} {Correlation functions in Fourier representation  with the effect of fluctuations  taken into account for $T^*=0.25$, $\bar\zeta=0.55$  (panel a) and for $T^*=0.2$, $\bar\zeta=0.55$ (panel b).  $\tilde{G}^*_{\alpha\beta}=\tilde{G}_{\alpha\beta}/\bar\zeta_{\alpha}\bar\zeta_{\beta}$, $\bar\zeta_{\alpha}=\pi\rho_{\alpha}\sigma_{\alpha}^3/6$, and the wave number $k$ is in $a^{-1}$ units with $a=(\sigma_{+}+\sigma_{-})/2$. The results are for the model with $\delta=0.5$,  $\epsilon_{++}^*=5$, $\epsilon_{++}^*=1$. In the insets, we show sharp peaks of $\tilde{G}^*_{--}(k)$.  }
		}
	\end{center}
\end{figure}

From ${\bf G}={\bf C}^{-1}$, one can calculate the correlation functions in Fourier representation. In Fig.~\ref{fig:Gij_fl}, we show the reduced correlation functions in Fourier  representation $\tilde{G}^*_{\alpha\beta}(k)=\tilde{G}_{\alpha\beta}(k)/\bar\zeta_{\alpha}\bar\zeta_{\beta}$ for the fixed total volume fraction $\bar\zeta=0.55$ and for two temperatures, $T^*=0.25$ (panel~a) and $T^*=0.2$ (panel~b). The three correlation functions $\tilde{G}^*_{\alpha\beta}(k)$ have sharp maxima for $k=k_{0}\simeq 1.58$. For both temperatures the  height  of the $\tilde{G}^*_{++}(k)$ maximum is about 100 times higher than the maximum  of $\tilde{G}^*_{--}(k)$. It should be noted that the dependence of $k_0$ on $T^*$ for the fixed $\bar\zeta$ is negligible, especially in the range $T^*=0.2-0.3$ (see Table~\ref{Table1}).
 \begin{table}
 	\centering
 	\caption{\label{Table1} The wave number $k_{0}$, the decay length $\alpha_{0}^{-1}$ and the period of oscillations $\lambda=2\pi/\alpha_{1}$ of 
 		the pair correlation functions $G_{\alpha\beta}(r)$ depending on the 
 		total number density $\bar\zeta$ for  fixed values of  temperature $T^{*}$. 
 		$T^{*}$  is defined in (\ref{red_units}), $\bar\zeta=\bar\zeta_{+}+\bar\zeta_{-}$,  $\bar\zeta_{\alpha}=\pi\rho_{\alpha}\sigma_{\alpha}^3/6$, $k_{0}$, $\alpha_{0}$ and $\alpha_{1}$ are in $a^{-1}$ units.}	
 	\vspace{2mm}
 	\begin{tabular}{| c | c | c | c | c | c | c| c } 
 		\hline
 \hspace{4mm}	 $T^{*}$\hspace{4mm} &  \hspace{4mm}  $\bar\zeta$ \hspace{4mm} &\hspace{4mm} $k_{0}$  \hspace{4mm}& \hspace{4mm} $\alpha_{0}$ \hspace{4mm}  &\hspace{4mm}  $\alpha_{1}$ \hspace{4mm} &  \hspace{4mm} $\alpha_{0}^{-1}$ \hspace{4mm} & \hspace{4mm}  $2\pi/\alpha_{1}$ \hspace{2mm} \\  
 		\hline
 		0.4  & 0.45 & 1.597 &0.225  &1.613  & 4.444  & 3.895 \\ \hline 
 		0.4  & 0.5  & 1.591 & 0.163& 1.599 & 6.151 &3.930 \\ \hline   
 		0.4  & 0.55 & 1.582 & 0.119 &1.587 & 8.424 & 3.959 \\ \hline  
 		0.3  & 0.45 & 1.602 & 0.124 &1.607 &8.042 & 3.910\\ \hline  
 		0.3  & 0.5  & 1.593 &0.088 &1.595  &11.426 &3.939 \\ \hline  
 		0.3  & 0.55 & 1.584 & 0.063 & 1.585 &15.796 &3.964 \\ \hline  
 		0.25 & 0.45 & 1.605 &0.084 &1.607& 11.852 & 3.911\\ \hline 
 		0.25 & 0.5  &1.594 &0.059 & 1.595 & 16.920&3.939\\ \hline 
 		0.25 & 0.55 &1.584 & 0.043 &1.585 & 23.411&3.964\\ \hline  
 		0.2  & 0.45 &1.607& 0.053 &1.608& 19.016 &3.907\\ \hline 
 		0.2  & 0.5  &1.595 &0.037 & 1.596 & 27.191& 3.937\\ \hline 
 		0.2  & 0.55 &1.585 & 0.027 &1.585 & 37.610& 3.963\\ 
 		\hline		
 	\end{tabular}
 	\centering
 \end{table}
 
\begin{figure}[htbp]
	\begin{center}
		\includegraphics[width=0.45\textwidth,angle=0,clip=true]{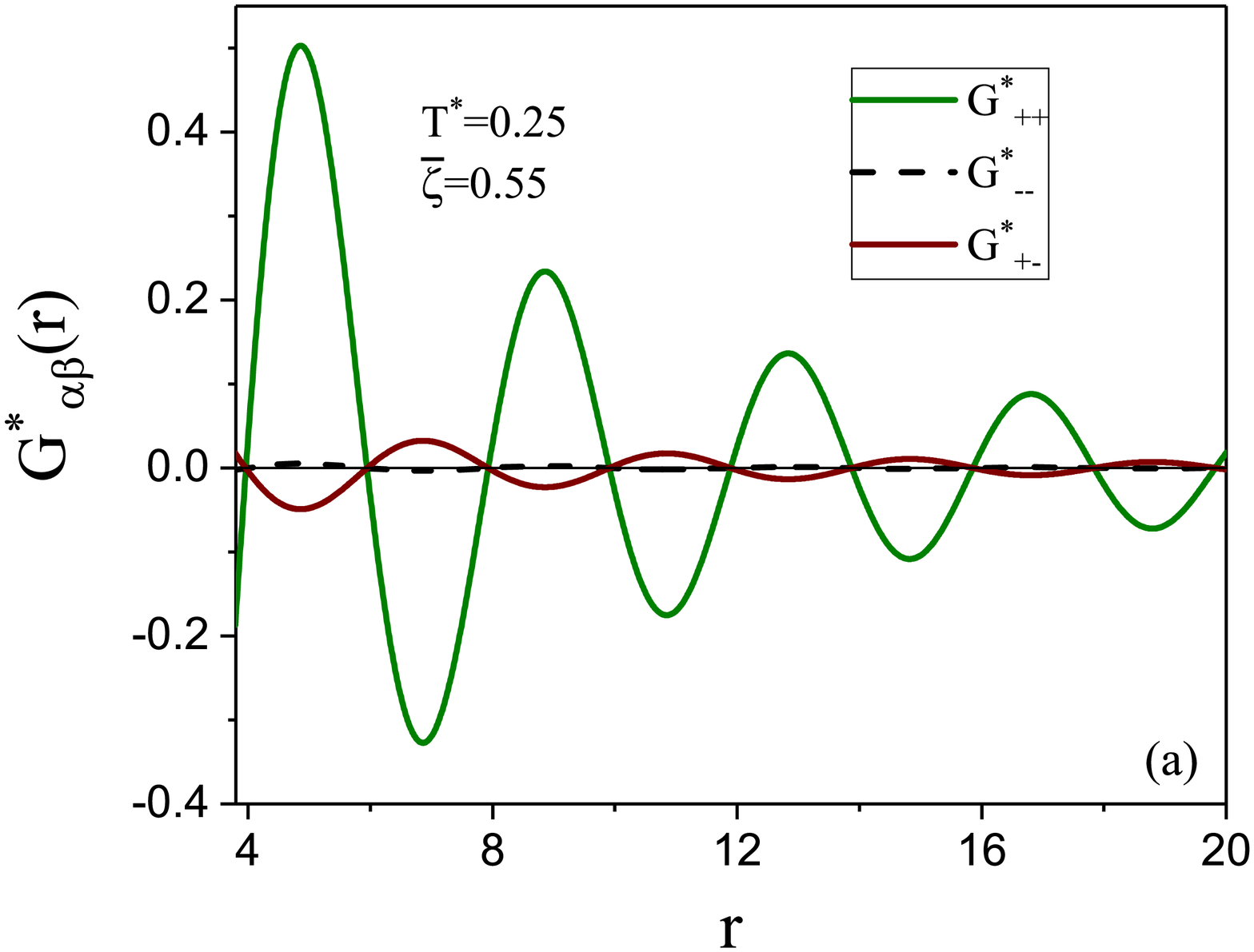}
		\includegraphics[width=0.46\textwidth,angle=0,clip=true]{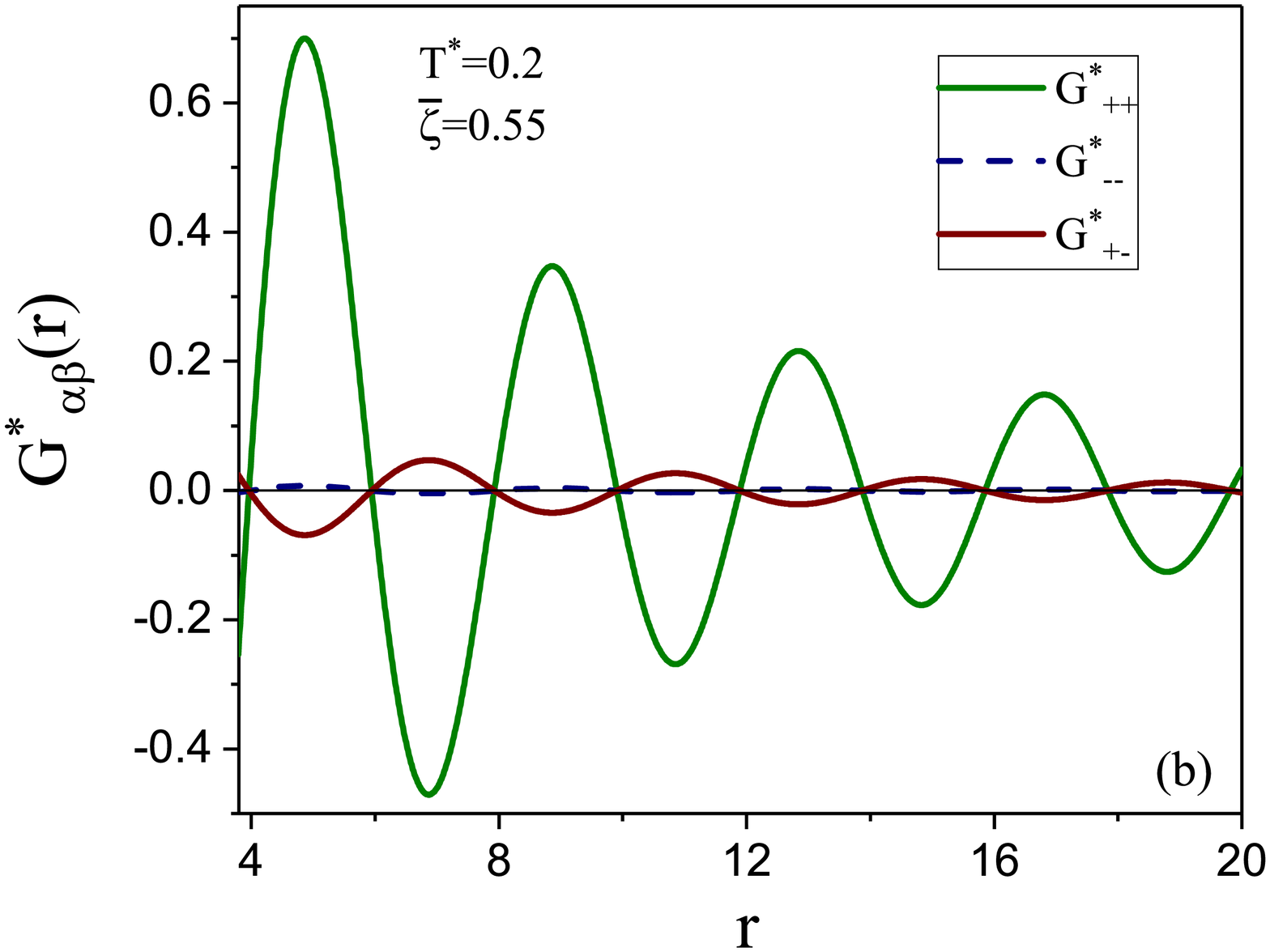}
		\caption{\label{fig:GIJ-r_fl} {Correlation functions  in real space  with the effect of fluctuations taken into account for $\bar\zeta=0.55$, $T^*=0.25$  (panel~a) and $T^*=0.2$ (panel~b). $G^*_{\alpha\beta}=G_{\alpha\beta}/\bar\zeta_{\alpha}\bar\zeta_{\beta}$, $\bar\zeta_{\alpha}=\pi\rho_{\alpha}\sigma_{\alpha}^3/6$, and  $r$ is in $a$ units with $a=(\sigma_{+}+\sigma_{-})/2$. The results are for the model with $\delta=0.5$,  $\epsilon_{++}^*=5$, $\epsilon_{++}^*=1$. }
		}
	\end{center}
\end{figure}

\begin{figure}[h]
 		\begin{center}
	\includegraphics[clip,width=0.45\textwidth,angle=0]{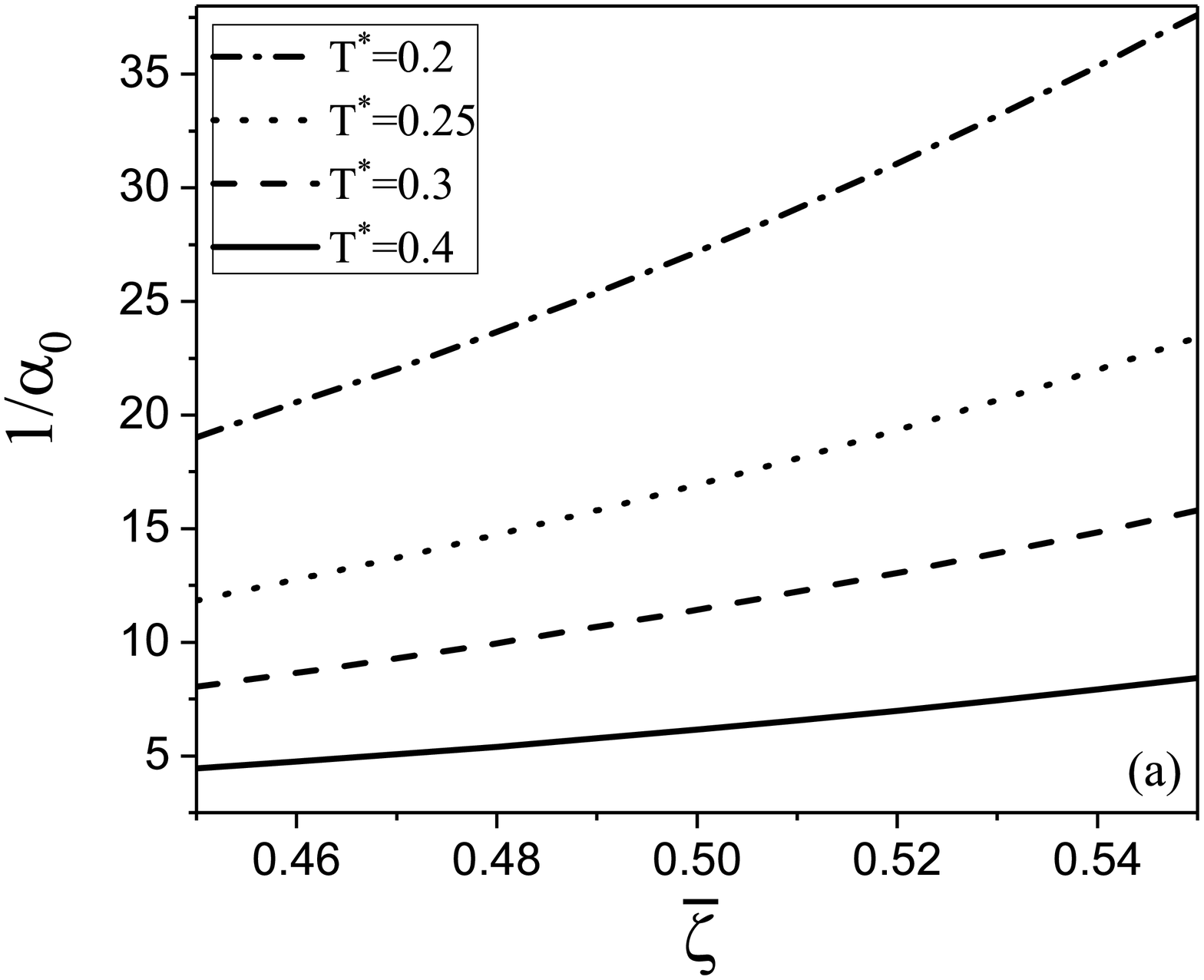}
 	\includegraphics[clip,width=0.47\textwidth,angle=0]{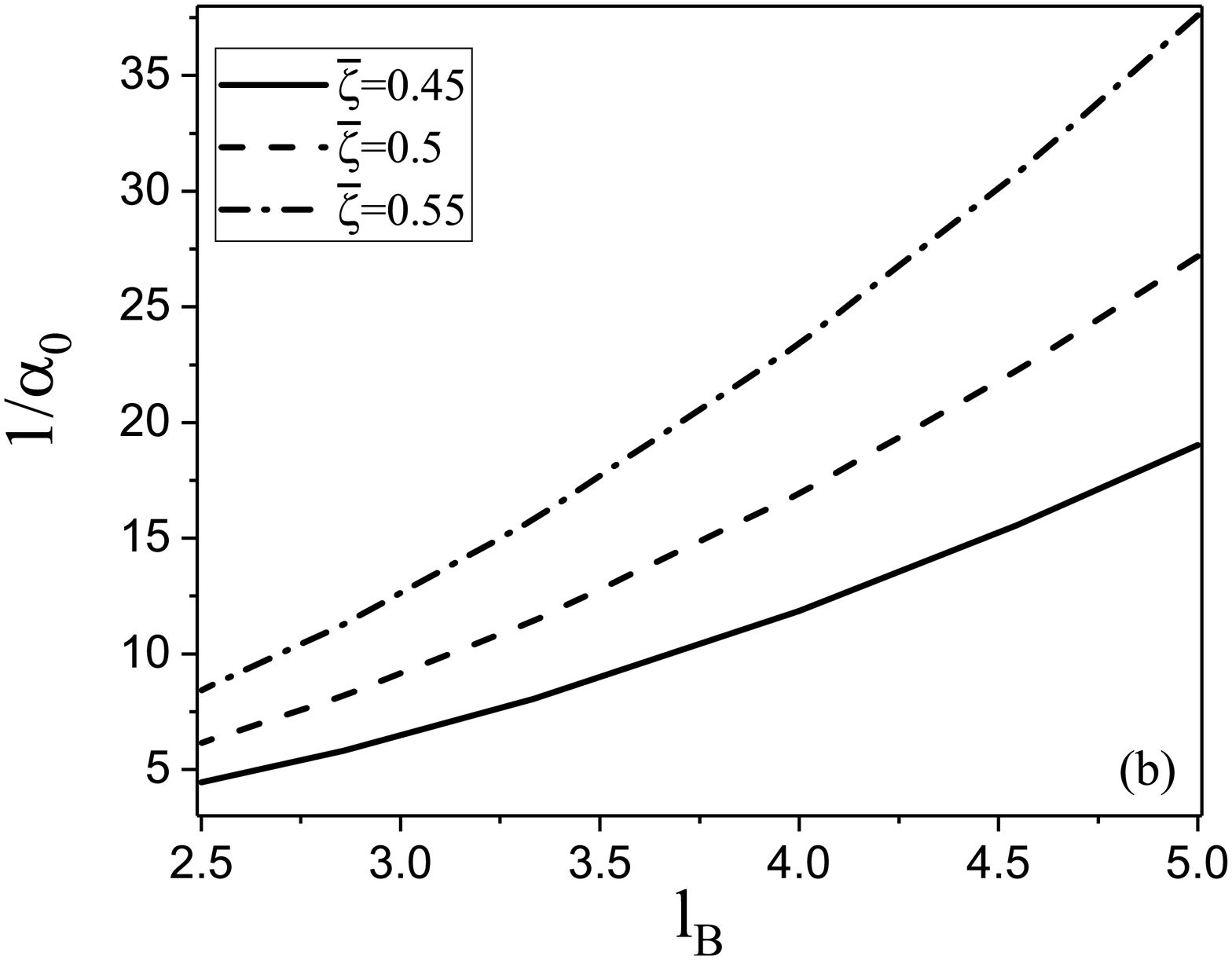}
 	\caption{\label{fig6:alpha0}
 	 The decay length $\alpha_{0}^{-1}$ of the correlation functions $G_{\alpha\beta}(r)$  as a function of the total volume fraction of ions $\bar\zeta$ for  $T^*=0.2$, $0.25$, $0.3$, $0.4$ (from the top to the bottom line) (panel a) and as a function of the Bjerrum length $l_{B}$ for $\bar\zeta=0.45$, $0.5$, $0.55$ (from the bottom to the top line) (panel b).
 	 $1/\alpha_{0}$ and $l_{b}$ are in the $a$ units, $a=(\sigma_{+}+\sigma_{-})/2\approx 0.4nm$. The results are for the model with $\delta=0.5$,  $\epsilon_{++}^*=5$, $\epsilon_{++}^*=1$.
 	}
 	\end{center}
 \end{figure}

  The reduced correlation functions  in real-space representation,  $G_{\alpha\beta}^*(r)=G_{\alpha\beta}(r)/\bar\zeta_{\alpha}\bar\zeta_{\beta}$, are obtained from
 the inverse Fourier transformation of  $\tilde{G}^*_{\alpha\beta}(k)$. They  are shown in Fig.\ref{fig:GIJ-r_fl} for  $T^*=0.25$, $\bar\zeta=0.55$ (panel~a) 
  and for $T^*=0.2$,  $\bar\zeta=0.55$ (panel~b). 
 As it is seen,   $G_{\alpha\beta}^*(r)$ show exponentially damped oscillatory behavior described by~(\ref{Gr}). 
 The period of damped oscillations is about $4a$. 
  We study the asymptotic decay of the correlation functions $G_{\alpha\beta}(r)$ using the pole analysis.  The  results
 of this numerical analysis for $T^*=0.2$, $0.25$, $0.3$, and $0.4$ and for $\bar\zeta=0.45, 0.5, 0.55$  are summarized in Table~\ref{Table1}. For the fixed volume fraction, the period $\lambda=2\pi/\alpha_{1}$  coincides with $2\pi/k_{0}$   and is rather kept  constant  for $T^*\leq 0.3$.
 For the fixed temperature,  $\lambda$ is a weakly increasing  function  of $\zeta$.
It should be noted that the period of damped oscillations obtained with the effect of fluctuations taken into account is  very close to the period obtained in MF for the higher temperature. By contrast, the decay length $\alpha_{0}^{-1}$ noticeably  increases with an increase of  $\bar\zeta$ for the fixed temperature and this increase is more rapid  for lower temperatures.  In Fig.~\ref{fig6:alpha0}, we present the  decay length $\alpha_{0}^{-1}$ as a function of the total volume fraction of ions $\bar\zeta$ for fixed  temperatures (panel~a) and as a function of the Bjerrum length $l_{B}$ for fixed volume fractions (panel~b). One can observe that  $\alpha_{0}^{-1}$ has a nearly linear dependence on the volume fraction for fixed $T^*$, with the slope decreasing with $T^*$ and a nearly linear dependence on $l_{B}$ for fixed $\bar\zeta$, with the slope increasing with $\bar\zeta$. 
 
 \section{Summary and conclusions}
 
 We have developed a highly simplified model for water-in-salt electrolytes, and focused on the salt LiTFSI that was a subject of recent experiments~\cite{Groves2021,borodin:17:0}. In our model, the ions are treated as charged hard spheres with different diameters, and we assumed additional, water-mediated specific interactions between the ions of the same sign. We assumed that  the solvent influences the distribution of the ions mainly by inducing effective interactions between them, and otherwise the solvent can be neglected.
 Next, we assumed that the detailed shape of the specific interactions is not important, as long as these interactions are strongly attractive but of a short range. We chose the Yukawa potentials for the specific interactions, and adjusted the parameters to  the  LiTFSI by requiring  the same scale of inhomogeneities 
 as found experimentally. Importantly, the obtained sum of the Coulomb and specific interactions for both, the anions and the cations 
 is attractive at short- and repulsive at large distances.

 For this model, we calculated correlation functions for concentrated electrolytes  for reduced temperatures close to the room temperature, using our theory for  binary mixtures with competing interactions~\cite{ciach:11:2,ciach:20:1,patsahan:21:0}. There is no unique way of relating the volume fraction in the approximate theory  to experimental molarity, since we assumed spherical rather than ellipsoidal anions, and the values of the ion diameters in the theory are not precise. We considered volume fractions $0.45\le \bar\zeta\le 0.55$ somewhat larger than in the experiment, but of the same order of magnitude (the molarity $0.38-0.46$ considered in~\cite{Groves2021} for spherical ions of our sizes gives $0.27\le \bar\zeta\le 0.32$).   We obtained exponentially damped oscillations for the correlation functions with the period $\lambda\approx 4$ in $a\approx 0.4nm$-units (in the experiment~\cite{Groves2021,borodin:17:0}, $\lambda\approx 1.4 nm$),  very weakly depending on $\bar\zeta$ in the considered range of concentrations. In MF,  $\lambda$  decreases a bit, whereas when the fluctuations are taken into account, it a bit increases with  $\bar\zeta$. Interestingly, $\lambda$ slightly decreases, increases or does not change with increasing $\bar\zeta$, depending on the method of determining it in the experiment~\cite{Groves2021}. 
 
 We found that the decay length of the correlations, $\lambda_s$,  increases almost linearly with $\bar\zeta$ for fixed reduced temperature $T^*$. It increases also with the Bjerrum length for temperatures of the order of the room temperature.  In particular, for the reduced temperature 
 $T^*=0.3$ (Bjerrum length $l_B=3.3$ in $a\approx 0.4nm$ units), we obtain 
 $\lambda_s\approx 3.2 - 6 nm$,
 and for 
 $T^*=0.2$ (Bjerrum length $l_B=5$) we obtain $\lambda_s\approx 8 - 15nm$ for $\bar\zeta=0.45 - 0.55$. These values are in semiquantitative agreement with the experimental decay lengths, $\lambda_s= 8.3  - 11.5nm$ for the molarity $3.8 - 4.6 M$ ($ \bar\zeta =0.27 - 0.32$).
 
 The very weak dependence of $\lambda$ on $\bar \zeta$ and the  semiquantitative agreement of the decay length with the experimental results indicate that the properties of the correlation functions do not depend sensitively on the details of the interactions. The ellipsoidal  anions are approximated by the spherical ones, implicit solvent inducing effective anion-anion and cation-cation interactions is assumed, and we neglected fluctuations of  the dielectric constant induced by the concentration fluctuations. The dependence of the reduced temperature on $\bar \zeta$ (through the dependence of $\epsilon $ on $\bar \zeta$) was disregarded as well. The latter dependence for the rather narrow range of $\bar\zeta$ is  not very strong, however. Finally, we rather arbitrarily assumed the shape of the specific interactions. With the above simplifications, we got semiquantitative  agreement with experiments. It means  that the above features are important only on the quantitative level. 
 
 We conclude that the key property determining the inhomogeneities on the mesoscopic length scale is the shape of the  sum of the Coulomb and the solvent-induced specific interactions. In order to induce mesoscopic inhomogeneities, this sum should be  attractive at short- and repulsive at large distances, with the ranges and strengths of the attractive and repulsive parts determined by the properties of the ions and the solvent.
 If these anion-anion and cation-cation  potentials have a negative minimum followed by a positive maximum, then layers of ions of the same sign of the thickness determined by the width and depth of the attractive well can be formed. We believe that this conclusion is not restricted to the particular case of the water-in-LiTFSI, but rather general. 
 
 Finally, we have shown that the self-consistent theory with the local fluctuations of the volume fractions taken into account~\cite{ciach:20:1,patsahan:21:0} can predict the structure with local inhomogeneities on a semiquantitative level. 
 
 There remains one unsolved problem - namely, the experimental disjoining pressure between crossed mica cylinders decays monotonically at large distances~\cite{Groves2021}, whereas our theory predicts the oscillatory decay. The same problem concerns simple salts and some other IL modeled by the RPM.

 \section{Appendices}

\subsection{ Interaction potentials in Fourier representation}
\label{AA}
The interaction potentials $\beta\tilde U_{\alpha\beta}^{C}$ and  $\beta\tilde U_{\alpha\alpha}^{A}$ (see (\ref{U++})--(\ref{Uii-r})) in Fourier representation   read
  \begin{eqnarray}
 \beta\tilde{U}_{++}^{C}(k)& =&\frac{4\pi}{T^*}\frac{\cos(k(1-\delta))}{k^2}, 
 \label{U++_C-k}
  \\
\beta\tilde{U}_{--}^{C}(k)& =&\frac{4\pi}{T^*}\frac{\cos(k(1+\delta))}{k^2}, 
\label{U--_C-k}
\\
 \beta\tilde{U}_{+-}^{C}(k)& =&-\frac{4\pi}{T^*}\frac{\cos(k)}{k^2},
 \label{U+-_C-k}
 \end{eqnarray}
 \begin{eqnarray}
 \beta\tilde U_{++}^{A}(k)&=&-\frac{4\pi \epsilon_{++}^{*}(1-\delta)}{T^*(z^2+k^2)}\Big[\cos(k(1-\delta))+\frac{z}{k}\sin(k(1-\delta))\Big],
 \label{U++_A-k}
 \\ 
 \beta\tilde U_{--}^{A}(k)&=&-\frac{4\pi \epsilon_{--}^{*}(1+\delta)}{T^*(z^2+k^2)}\Big[\cos(k(1+\delta))+\frac{z}{k}\sin(k(1+\delta))\Big],
 \label{U--_A-k}
 \end{eqnarray}
where  $z$ and $k$ are in $a^{-1}$ units.
\subsection{Free-energy density for a mixture of hard spheres with different diameters}
For the free energy density $\beta f_{hs}$ we use the expression obtained in the Carnahan-Starling approximation \cite{Mansoori:71}
\begin{eqnarray*}
\beta f_{hs}=\frac{6}{\pi(1-\delta^2)^3}[\zeta_{+}(1+\delta)^3+\zeta_{-}(1-\delta)^3]\left[-\frac{3}{2}(1-y_{1}+y_{2}+y_{3})
\right.
\nonumber \\
\left.
+\frac{3y_{2}+2y_{3}}{1-\zeta}+\frac{3}{2}\frac{(1-y_{1}-y_{2}-y_{3}/3)}{(1-\zeta)^{2}}+(y_{3}-1)\ln(1-\zeta)\right], 
\end{eqnarray*}
with
\begin{eqnarray*}
y_{1}&=&\frac{8\delta^2\zeta_{+}\zeta_{-}}{\zeta[\zeta_{+}(1+\delta)^3+\zeta_{-}(1-\delta)^3]},\nonumber \\
y_{2}&=&\frac{4\delta^2\zeta_{+}\zeta_{-}}{\zeta^2}\frac{[\zeta_{+}(1+\delta)+\zeta_{-}(1-\delta)]}{[\zeta_{+}(1+\delta)^3+\zeta_{-}(1-\delta)^3]},\nonumber \\
y_{3}&=&\frac{[\zeta_{+}(1+\delta)+\zeta_{-}(1-\delta)]^3}{\zeta^2[\zeta_{+}(1+\delta)^3+\zeta_{-}(1-\delta)^3]},
\end{eqnarray*}

\subsection{Expressions for $\tilde {\bf C}^{MF}(k)$}
 Taking into account (\ref{U++_C-k})-(\ref{U--_A-k}), (\ref{V}), and (\ref{A4}), 
 the matrix ${\bf \tilde{C}}^{MF}$ can be presented as follows:
\begin{eqnarray}
\tilde C_{++}^{MF}(k)&=&\frac{144 \cos(k(1-\delta))}{T^*\pi(1-\delta)^6k^2}-\frac{144\epsilon^{*}_{++}}{T^*\pi(1-\delta)^5(z^2+k^2)}\Big[\cos(k(1-\delta))\nonumber \\
&&+\frac{z}{k}\sin(k(1-\delta))\Big]+A_{++},
\label{C++_mf} \\
\tilde C_{--}^{MF}(k)&=&\frac{144 \cos(k(1+\delta))}{T^*\pi(1+\delta)^6k^2}-\frac{144\epsilon^{*}_{--}}{T^*\pi(1+\delta)^5(z^2+k^2)}\Big[\cos(k(1+\delta)) \nonumber \\
&&+\frac{z}{k}\sin(k(1+\delta))\Big]
+A_{--},
\label{C--_mf} \\
\tilde C_{+-}^{MF}(k)&=&-\frac{144\cos(k)}{T^*\pi(1-\delta^2)^3k^2}+A_{+-},
\label{C+-_mf}
\end{eqnarray}
where 
\begin{equation}
A_{\alpha\beta}=\frac{6}{\pi}\frac{\delta_{\alpha\beta}}{\zeta_{\alpha} a_{\alpha}^3}+\frac{\partial^{2}\beta f_{hs}}{\partial\zeta_{\alpha}\partial\zeta_{\beta}},
 \qquad
 a_{\pm}=1\mp\delta.
\label{Aij}
\end{equation} 
\subsection{Equations for $k_0$ and $\tilde C_{\alpha\beta}(k_0)$ in our mesoscopic theory.}

The equations to be solved for $k_0$ and $\tilde C_{\alpha\beta}(k_0)$ are (see sec.\ref{theory} and Ref.\cite{ciach:20:1})
 \begin{equation*}
 \tilde V_{++}^{'}(k_0)\tilde C_{--}(k_0)+\tilde V_{--}^{'}(k_0)\tilde C_{++}(k_0)
 -2\tilde V_{+-}^{'}(k_0)\tilde C_{+-}(k_0)=0,
 \end{equation*}
  \begin{eqnarray*}
 \tilde C_{++}(k_0)&=&\tilde C_{++}^{MF}(k_0) +
 \frac{k_0^2}{2\pi[2\beta \tilde W^{''}(k_0)D_0]^{1/2}}\left[A_{++++}\tilde C_{--}(k_0)-2A_{+++-}\tilde C_{+-}(k_0) \right. \nonumber \\
 && \left.
 +A_{++--}\tilde C_{++}(k_0)\right], 
 \\
 \tilde C_{--}(k_0)&=&\tilde C_{--}^{MF}(k_0) +
 \frac{k_0^2}{2\pi[2\beta \tilde W^{''}(k_0)D_0]^{1/2}}\left[A_{++--}\tilde C_{--}(k_0)-2A_{+---}\tilde C_{+-}(k_0) \right. \nonumber \\
 && \left.
 +A_{----}\tilde C_{++}(k_0)\right], 
  \\
 \tilde C_{+-}(k_0)&=&\tilde C_{+-}^{MF}(k_0)+
 \frac{k_0^2}{2\pi[2\beta \tilde W^{''}(k_0)D_0]^{1/2}}\left[A_{+++-}\tilde C_{--}(k_0)-2A_{++--}\tilde C_{+-}(k_0)\right. \nonumber \\ 
 &&\left.
 +A_{+---}\tilde C_{++}(k_0)\right].
 \end{eqnarray*}
where
 \begin{eqnarray}
\beta W^{''}(k_0)=2 \beta^2 \det {\bf V}^{'}+\beta \tilde V_{++}^{''}\tilde C_{--}(k_0)+\beta \tilde V_{--}^{''}\tilde C_{++}(k_0)-2\beta \tilde V_{+-}^{''}\tilde C_{+-}(k_0),
\label{W''}
\end{eqnarray}
 with
 $\tilde{V}_{\alpha\beta}^{'}(k_0)=\rm{d}\tilde V_{\alpha\beta}(k)/\rm{d}k|_{k=k_0}$, $\tilde{C}_{\alpha\beta}^{MF}(k_0)$ are obtained from (\ref{C++_mf})- (\ref{Aij}) by putting $k=k_{0}$,  $D_0$ is presented in (\ref{D0}), and 
 \begin{equation}
 \label{Aijkl}
 A_{\alpha\beta\gamma\delta}=\frac{12}{\pi}\frac{\delta_{\alpha\beta}\delta_{\alpha\gamma}\delta_{\alpha\delta}}{\zeta_{\alpha}^3 a_{\alpha}^3}+\frac{\partial^{4}\beta f_{hs}}{\partial\zeta_{\alpha}\partial\zeta_{\beta}\partial\zeta_{\gamma}\partial\zeta_{\delta}}.
 \end{equation}
\bibliography{bibliography_20}

\end{document}